\documentclass[graybox]{svmult}
\usepackage{mathptmx}       
\usepackage{graphicx}        
\usepackage[bottom]{footmisc}

\usepackage{amsfonts}
\usepackage{amssymb}
\usepackage{textcomp}
\usepackage[colorlinks=true,citecolor=blue,urlcolor=blue]{hyperref}
\newcommand{\bea}{\begin{eqnarray}}
\newcommand{\eea}{\end{eqnarray}}
\newcommand{\beann}{\begin{eqnarray*}}
\newcommand{\eeann}{\end{eqnarray*}}
\newcommand{\be}{\begin{equation}}
\newcommand{\ee}{\end{equation}}
\newcommand{\benn}{\begin{equation*}}
\newcommand{\eenn}{\end{equation*}}
\newcommand{\rmL}{{\rm L}}
\newcommand{\rmR}{{\rm R}}
\newcommand{\rmC}{{\rm C}}
\newcommand{\kBT}{k_{\rm B}T}
\newcommand{\nB}{n_{\rm B}}

\begin{document}

\title{Transport out of locally broken detailed balance}
\author{Rafael S\'anchez}
\institute{Instituto Gregorio Mill\'an, Universidad Carlos III de Madrid, 28911 Legan\'es, Madrid, Spain
\at Departamento de F\'sica Te\'orica de la Materia Condensada, Universidad Aut\'onoma de Madrid, 28049 Madrid, Spain
}
%
%
\maketitle

\abstract*{Each chapter should be preceded by an abstract (10--15 lines long) that summarizes the content. The abstract will appear \textit{online} at \url{www.SpringerLink.com} and be available with unrestricted access. This allows unregistered users to read the abstract as a teaser for the complete chapter. As a general rule the abstracts will not appear in the printed version of your book unless it is the style of your particular book or that of the series to which your book belongs.
Please use the 'starred' version of the new Springer \texttt{abstract} command for typesetting the text of the online abstracts (cf. source file of this chapter template \texttt{abstract}) and include them with the source files of your manuscript. Use the plain \texttt{abstract} command if the abstract is also to appear in the printed version of the book.}

\abstract{Electrons move along potential or thermal gradients. In the presence of a global gradient, applied e.g. to the two terminals of a conductor, this induces electric charge and heat currents. They can also flow between two equilibrated terminals (at the same voltage and temperature) if detailed balance is broken in some part of the system. A minimal model involving two metallic islands in series is introduced whose internal potential and temperatures can be externally modulated. The conditions for a finite electric flow are discussed. }

\section{Introduction}
An electronic conductor responds to a nonequilibrium situation in the form of charge and heat currents. It can be due to the presence of electric or thermal gradients applied to the two terminals of the system, $V_\rmL-V_\rmR$ and $T_\rmL-T_\rmR$. Transport is however not restricted to this situation. The role of fluctuations in the intermediate region (the system) cannot be overlooked. If the two terminals are in thermal equilibrium (for being at the same temperature and potential) a current can nevertheless flow due to rectified noise.  

\begin{figure}[t]
\sidecaption
\includegraphics[width=2.9in,clip]{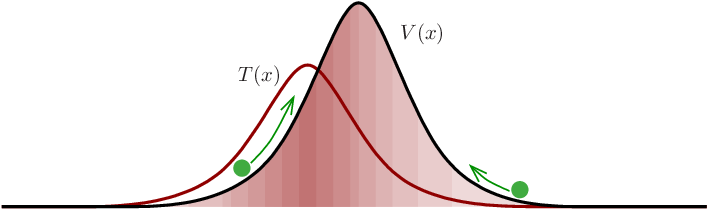}
\caption{\label{barrier}\small  Transport in the presence of state-dependent noise. Symmetric potential and temperature distributions give nevertheless a finite particle current if their respective maxima are displaced~\cite{markus,vankampen,landauer}.}
\end{figure}

The relevance of noisy states was pointed out by Landauer for modifying the relative occupation of bistable potentials~\cite{blowtorch}. This is the case for instance if temperature is locally increased (with an ideal blowtorch) on one side of the barrier only~\cite{blowtorch}. These ideas were later applied to transport configurations in a series of papers by B\"uttiker~\cite{markus}, van Kampen~\cite{vankampen} and Landauer~\cite{landauer}. They predict classical particles to flow along symmetric potentials which are subject to state-dependent noise. As an example, particles overcome a potential barrier from the side where the diffusion coefficient is higher (due e.g. to a locally increased temperature, cf. Fig.~\ref{barrier}), leading to a broken detailed balance situation. The resulting currents depend on the separation of the potential and temperature fields: symmetric state-dependent potential and temperature distributions give no transport~\cite{markus}. Of course Onsager relations are still fulfilled~\cite{vankampen_Onsager_1991} if one takes into account the heat injected (by the blowtorch) in order to keep the temperature stationary.

Similar ideas have motivated proposals of thermal ratchets~\cite{yaroslav,benjamin} and thermoelectric metamaterials~\cite{Humphrey_Reversible_2005}, and the measurement of radiation induced currents~\cite{olbrich,olbrich2} or transverse rectification in semiconductor 2DEGs~\cite{ganczarczyk,rojek}. They are closely related to the recent field of mesoscopic three-terminal thermoelectrics~\cite{review,benenti_fundamental_2016} which discusses the conversion of a heat current injected from a thermal reservoir into a charge current in an equilibrium conductor~\cite{hotspots,holger,comptes,entin_three_2010,ruokola_theory_2012,
bjorn,roche_harvesting_2015,jiang_thermoelectric_2012,jiang_three_2013,jiang_hopping_2013,
mazza_thermoelectric_2014,entin_enhanced_2015,mazza_separation_2015,bosisio_nanowire_2016}. Also noise generated either in a coupled conductor~\cite{drag,bischoff_measurement_2015,keller_cotunneling_2016} or externally~\cite{hartmann_voltage_2015,pfeffer_logical_2015} is used for this purpose. Most of these cases use discrete levels in quantum dot systems where a temperature cannot be properly defined~\cite{rob}. Another possibility is a mesoscopic thermocouple configuration~\cite{Humphrey_Reversible_2002,swisscheese,wells,chiral,whitney_quantum_2016} where the current is due to the separation of electron-hole exitacions at the two contacts of a hot cavity. Common to all of them is the need of a conductor with broken left-right inversion and electron-hole symmetries (usually due to energy-dependent transport coefficients).

\begin{figure}[t]
\sidecaption
\includegraphics[width=2.5in,clip]{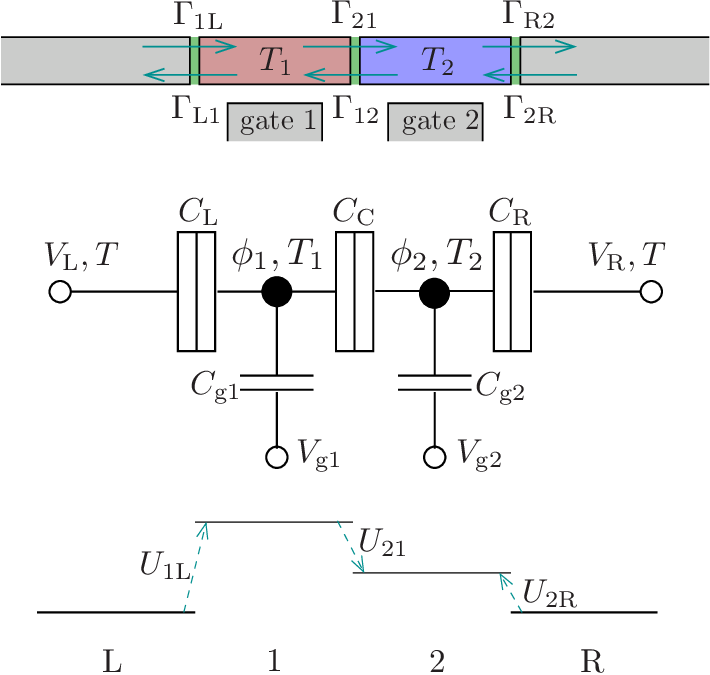}
\caption{\label{sys}\small  Sketch of our system. Two metallic islands are tunnel-coupled in series in a two-terminal configuration. Electrons are assumed to thermalise at a different temperature $T_i$ in each piece of the conductor. The island internal potentials can be tuned by gate voltages $V_{{\rm g},i}$. The two terminals are considered at the same voltage, $V_{\rm L}=V_{\rm R}=0$, and temperature, $T$. The piecewise Coulomb gap $U_{ji}$ experienced by tunneling electrons can be modulated with the gate voltages $V_{{\rm g}k}$.}
\end{figure}

Here a minimal model of a mesoscopic conductor is proposed where all the necessary ingredients can be engineered and controlled experimentally. It is based on the discretization of a metallic conductor by means of at least three tunnel junctions, cf. Fig.~\ref{sys}. This way an array of two small metallic islands are formed whose level spacing is much smaller than the thermal energy, $\Delta E\ll\kBT$. The electron-electron relaxation rate is fast such that they equilibrate to a Fermi distribution function. Hence, each island has a well defined temperature, $T_i$. A physical {\it mesoscopic blowtorch} can be defined that modulates the temperature of the system locally. This can be done either by introducing time dependent drivings~\cite{averin_statistics_2011,pekola,saira}, or by using either on-chip refrigerators~\cite{nahum_electronic_1994,leivo,timofeev,koski_on-chip_2015,partanen_quantum_2016,sinis} or the noise generated in a coupled conductor~\cite{kubala,drag}. In the first case, the frequency of the driving must be much larger than the energy relaxation rate such that electrons have an increased effective temperature in the island. In the second case, it is important that the refrigerator exchanges heat but no charge with the conductor~\cite{hotspots}. Additionally, the internal potential of each island can be externally modulated by means of gate voltages. This way, both temperature and potential profiles can be spatially resolved and tuned in a simple configuration. 

\section{Two Coulomb islands in series}

Our model is based on the well known physics of single-electron tunneling at small-capacitance tunnel junctions~\cite{averin}. Metallic islands can be separated by insulating barriers such that the energy cost for adding an extra electron (the charging energy, $E_{i}$) is larger than the thermal energy $E_{i}\gg\kBT$. For islands of a micrometer size, the charging energy is of the order of 0.1~meV. This introduces electron-electron correlations that suppress the low voltage transport, what is called Coulomb blockade. The extra energy can be supplied by a gate voltage $V_{{\rm g}i}$ such that electrons can flow one by one through the island giving rise to conductance~\cite{glazman,beenakker,sct} and thermopower~\cite{beenakker_theory_1992,staring_coulomb_1993,dzurak_observation_1993, moller_charging_1998,turek} oscillations. Electronic cooling based on these properties has been recently proposed~\cite{coulombgap} and realized experimentally~\cite{anna} in single-electron transistors.

Consider an array of three tunnel junctions forming two islands~\cite{pothier,pothier1992,lotkhov,limbach}, as sketched in Fig.~\ref{sys}. Each barrier is described by a capacitance $C_i$ and a resistance $R_i$ in parallel. The former determine the electrostatic potential of each island, $\phi_i$, which is obtained self-consistently from the equations for the charge accumulated in each island: 
\begin{eqnarray}
Q_1&=&C_\rmL(\phi_1-V_\rmL)+C_\rmC(\phi_1-\phi_2)+C_{{\rm g}1}V_{{\rm g}1}\\
Q_2&=&C_\rmR(\phi_2-V_\rmR)+C_\rmC(\phi_2-\phi_1)+C_{{\rm g}2}V_{{\rm g}2}.
\end{eqnarray} 
The electrostatic potential in the coupled system is $U(Q_1,Q_2)=\sum_i\int dQ_i\phi_i$. It defines the charging energies $E_i=e^2/2\tilde C_i$, with $\tilde C_i=\left(C_{\Sigma_1}C_{\Sigma2}-C_\rmC^2\right)/C_{\Sigma\bar{i}}$ (where $\bar{i}$ denotes the island next to $i$), in terms of the total geometric capacitance $C_{\Sigma i}=C_{\alpha_i}+C_\rmC+C_{{\rm g}i}$ of each island $i$ (with $\alpha_1=\rmL$ and $\alpha_2=\rmR$), the centre capacitance $C_\rmC$, and the one coupling to the gate, $C_{{\rm g}i}$. There is a crossed charging energy related to the occupation of the next island, parametrized as $J=2e^2C_\rmC/\left(C_{\Sigma_1}C_{\Sigma2}-C_\rmC^2\right)$. It corresponds to the repulsion of electrons in different islands. The internal potential of the islands is furthermore tuned by the control parameters 
\be
n_{{\rm g}i}=\frac{1}{e}\frac{4E_1E_2}{J^2-4E_1E_2}\left(Q_i^{\rm ext}-\frac{J}{4E_i}Q_{\bar{i}}^{\rm ext}\right),
\ee
with the externally induced charges~\cite{beenakker} $Q_1^{\rm ext}=C_\rmL V_\rmL+C_{{\rm g}1}V_{{\rm g}1}$ and $Q_2^{\rm ext}=C_\rmR V_\rmR+C_{{\rm g}2}V_{{\rm g}2}$. Here we are interested in the current generated in an unbiased configuration, such that $V_\rmL=V_\rmR=0$. Hence, $n_{{\rm g}i}$ depends only on the gate voltages. Note that each gate affects the two islands. A time-dependent modulation of the gates has been used for single-electron pumping with metrological precision~\cite{pothier1992,lotkhov} and provides a feed-back control mechanism for Maxwell's demon proposals~\cite{averin_maxwell_2011}.

All these can be cast into the electrostatic term of the Hamiltonian of the system:
\be
H=\sum_i E_i(n_i-n_{{\rm g}i})^2+J(n_1-n_{{\rm g}1})(n_2-n_{{\rm g}2}),
\ee
where $n_i=Q_i/e$ is the number of electrons in each island. Let us restrict here to the simplest configuration with up to one extra electron in the system. This is the case at low temperatures, $\kBT\ll E_i,J$, which for typical experimental situations is around 100~mK. The charge configuration of the system is then described by the three states $(n_1,n_2)$: $(0,0)$, $(1,0)$ and $(0,1)$. 

Tunneling events are charaterized by the energy cost $U_{ji}$ (the Coulomb gap) for electrons tunneling from region $i$ to $j$, which read:
\begin{eqnarray}
U_{1\rmL}&=&2E_1\left(\frac{1}{2}-n_{{\rm g}1}\right)-Jn_{{\rm g}2}\\
U_{2\rmR}&=&2E_2\left(\frac{1}{2}-n_{{\rm g}2}\right)-Jn_{{\rm g}1}\\
U_{21}&=&U_{2\rmR}-U_{1\rmL}
\end{eqnarray}
in the absence of a bias voltage. Obviously, $U_{ij}=-U_{ji}$. We emphasize that the energetics of the mesoscopic junction is fully tunable with the gate voltages. The rates for the corresponding tunneling transitions are given by the well-known expression:
\be
\label{rates}
\Gamma_{ji}=\frac{1}{e^2R_{ji}}\int dEf(E,T_i)[1-f(E-U_{ji},T_j)],
\ee
where $R_{ji}=R_\rmL$, $R_\rmC$ or $R_\rmR$, is the tunneling resistance of the involved junction. They are quite generally energy-independent (typically around 10-100~k$\Omega$). Here $f(E,T)=1/\left(1+e^{E/\kBT}\right)$ is the Fermi-Dirac distribution for a region at temperature $T$. 

With these, one can write the rate equations:
\bea
\dot{ P}_{(0,0)}&=&\Gamma_{\rmL1}P_{(1,0)}+\Gamma_{\rmR2}P_{(0,1)}-\left(\Gamma_{1\rmL}+\Gamma_{2\rmR}\right)P_{(0,0)}\nonumber\\
\dot P_{(1,0)}&=&\Gamma_{1\rmL}P_{(0,0)}+\Gamma_{12}P_{(0,1)}-\left(\Gamma_{\rmL1}+\Gamma_{21}\right)P_{(1,0)}\\
\dot P_{(0,1)}&=&\Gamma_{21}P_{(1,0)}+\Gamma_{2\rmR}P_{(0,0)}-\left(\Gamma_{12}+\Gamma_{\rmR2}\right)P_{(0,1)}\nonumber,
\eea
for the dynamics of the occupation probability of each state, $P_m(t)$. 
The stationary occupation of the different states are obtained by solving $\dot P=0$, giving:
\bea
\label{stat}
P_{(0,0)}&=&\Gamma_{\rm T}^{-2}\left(\Gamma_{21}\Gamma_{\rmR2}+\Gamma_{\rmL1}\Gamma_{12}+\Gamma_{\rmL1}\Gamma_{\rmR2}\right)\nonumber\\
P_{(1,0)}&=&\Gamma_{\rm T}^{-2}\left(\Gamma_{1\rmL}\Gamma_{\rmR2}+\Gamma_{12}\Gamma_{1\rmL}+\Gamma_{12}\Gamma_{2\rmR}\right)\\
P_{(0,1)}&=&\Gamma_{\rm T}^{-2}\left(\Gamma_{\rmL1}\Gamma_{2\rmR}+\Gamma_{21}\Gamma_{1\rmL}+\Gamma_{21}\Gamma_{2\rmR}\right),\nonumber
\eea
with $\Gamma_{\rm T}^2$ given by the normalization condition, $\sum_m P_m=1$. The stationary state obeys detailed balance if tunneling transitions satisfy: $\Gamma_{1\rmL}P_{(0,0)}=\Gamma_{\rmL1}P_{(1,0)}$, for the left $\Gamma_{12}P_{(1,0)}=\Gamma_{21}P_{(0,1)}$, for the center, and $\Gamma_{1\rmR}P_{(0,0)}=\Gamma_{\rmR1}P_{(0,1)}$, for the right junction. Hence, the current in the right terminal:
\be
I_\rmR=e\left(\Gamma_{R2}P_{(0,1)}-\Gamma_{2\rmR}P_{(0,0)}\right),
\ee
gives a measure of the deviation from detailed balance for processes at the right barrier. Injected currents from either terminal are defined as positive. Using Eqs.~(\ref{stat}) results in the expression:
\be
\label{eq:curr}
I_\rmR=e\Gamma_{\rm T}^{-2}\left(\Gamma_{1\rmL}\Gamma_{21}\Gamma_{\rmR2}-\Gamma_{\rmL1}\Gamma_{12}\Gamma_{2\rmR}\right).
\ee
From charge conservation, we obtain the current through the left junction: $I_\rmL=-I_\rmR$.
Note that the first term in the right hand side of Eq.~(\ref{eq:curr}) is proportional to the rate for an electron to be transported from the left to the right terminal (via a sequence L$\rightarrow$1$\rightarrow$2$\rightarrow$R). The second term is proportional to the rate of the reversed process. It is then clear that the case of having transitions satisfying local detailed balance at every junction results in no net current: In the unbiased and isothermal configuration, this translates to having tunneling rates related by $\Gamma_{ij}=\Gamma_{ji}e^{U_{ij}/\kBT}$. This is not the case if one of the leads or islands is at a different temperature, as discussed in the next section.

\subsection{Broken detailed balance. Linear response}
Let us first emphasize the importance of the Coulomb gap introduced by the electronic confinement in the island. If $U_{ji}=0$, tunneling is electron-hole symmetric and a temperature drop across the junction is not sufficient to break detailed balance. It can be easily shown from symmetries of the Fermi function that the integral
\be
\int_{-\infty}^{\infty} dE f(E,T)\left[1-f(E,T')\right]
\ee
is invariant under the exchange of temperatures $T\leftrightarrow T'$.
Hence $\Gamma_{ji}=\Gamma_{ij}$, for $U_{ji}=0$, independently of the temperatures $T_i$ and $T_j$. 

In the following, the case with finite $U_{ji}$ is assumed, unless explicitely mentioned. Some of the rates Eq.~(\ref{rates}) can be analytically calculated by using the relation:
\be
f(E,T)[1-f(E',T')]={\nB}{\left(\frac{E}{\kBT}{-}\frac{E'}{\kBT'}\right)}[f(E',T')-f(E,T)],
\ee
where $\nB(x)=\left(e^{x}-1\right)^{-1}$ is the Bose-Einstein distribution function. For transitions between regions at the same temperature one gets:
\be
\Gamma_{ji}^{(0)}=-\frac{U_{ji}}{e^2R_{ji}}\nB\left(\frac{U_{ji}}{\kBT}\right)e^{U_{ji}/\kBT}
\ee
and, on the other hand, using $\nB(-x)=-e^x\nB(x)$:
\be
\Gamma_{ij}^{(0)}=-\frac{U_{ji}}{e^2R_{ji}}\nB\left(\frac{U_{ji}}{\kBT}\right),
\ee
for the reversed process. It is straightforward to check that local detailed balance is satisfied at such a junction. If furthermore the gates are tuned such that there is no energy cost, $U_{ji}=0$, tunneling is governed by thermal fluctuations: $\Gamma_{ji}^{(0)}=\Gamma_{ij}^{(0)}=\kBT/(e^2R_{ji})$.

\begin{figure}[t]
\sidecaption
\includegraphics[width=1.7in,clip]{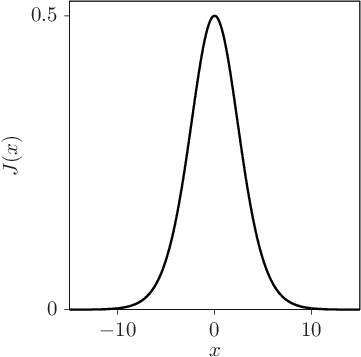}
\caption{\label{jx}\small  Function $J(x)$ defined in Eq.~(\ref{eq:jx}).}
\end{figure}

This is not the case when the junction separates two pieces of the metal which are at different temperatures. This effect shows up in the linear regime. Considering a small temperature difference $\delta T=T_j-T_i$, one can linearize the involved tunneling rates, $\Gamma_{ji}\approx\Gamma_{ji}^{(0)}+\Gamma_{ji}^{(1)}\delta T/T$, with $\Gamma_{ji}^{(1)}=\kBT(e^2R_{ji})^{-1}J(U_{ji}/\kBT)$ and:
\be
\label{eq:jx}
J(x)={e^{x}}\left[\nB\left(x\right)\right]^2\left[\frac{\pi^2}{6}+x^2+{\rm Li}_2\left(-e^{x}\right)+{\rm Li}_2\left(-e^{-x}\right)\right],
\ee
with the dilogarithm function ${\rm Li}_2(z)=\sum_{k=1}^\infty z^k/k^2$. The function $J(x)$ is a peak centered at the origin, as shown in Fig.~\ref{jx}. That is, the respose will be larger close to the Fermi energy.
From the above expression, we can verify that $J(x)=J(-x)$ and therefore $\Gamma_{ij}^{(1)}=\Gamma_{ji}^{(1)}$. Hence the rates do not satisfy detailed balance:
\be
\frac{\Gamma_{ji}}{\Gamma_{ij}}\approx e^{U_{ji}/\kBT}+\frac{\delta T}{T}F(U_{ji}/\kBT),
\ee
where $F(x)=x^{-1}J(x)[\nB(x)]^{-2}$. Note that the effect appears in the simultaneous occurance of a temperature drop and a Coulomb gap.

Using the above relations in the expression for the charge current, Eq.~(\ref{eq:curr}), one finds that breaking detailed balance in a single barrier is enough to have a finite current, even if the two terminals are at the same voltage and temperature. 

To be more specific, let us consider the case where only the first island is at a different temperature, $T_1=T+\delta T$, with $T_2=T$. Then a current is generated:
\be
I_\rmR\propto\frac{\delta T}{T}\left[e^{-U_{21}/\kBT}F(U_{21}/\kBT)-e^{U_{1\rmL}/\kBT}F(-U_{1\rmL}/\kBT)\right],
\ee
in terms of the energy cost for tunneling to the hot island from the left lead and from the other island, $U_{1\rmL}$ and $-U_{21}$, respectively. The avoided prefactor depends on the equilibrium rates, $\Gamma_{ji}^{(0)}$. It is clear from the above expression that no current will flow if detailed balance is broken symmetrically at the two barriers of the hot island, i.e. if the energy cost is equal: $U_{1\rmL}=-U_{21}$. This way, tuning the gate voltages allows one to control the current.

\section{Transport from a hotspot}

As discussed in the previous section, broken detailed balance occurs in tunneling through a junction separating regions at different temperatures. This is however not a sufficient condicion to generate transport in a conductor in global equlibrium. For example, no current will occur in a system consisting on a single hot island, as broken detailed balance is symmetric in the two junctions\footnote{Remember that tunneling resistances are energy independent in metalic islands. This is not necessarily the case in other related systems with energy-dependent tunneling junctions, e.g. in semiconductor quantum dots.}.
In order to have a current, an asymmetry needs to be introduced, e.g. by making the energy cost for tunneling through the two barriers different. In our case, this is the role of the second island. Its charging energy lifts the asymmetry making $U_{1\rmL}$ and $U_{12}$ different. Furthermore, this asymmetry can be tuned with gate voltages $V_{{\rm g}1}$ and $V_{{\rm g}2}$, as discussed above. If the second island is at the same temperature than the two terminals, $T_2=T$, detailed balance is satisfied at the third junction, with $\Gamma_{2\rmR}=\Gamma_{\rmR2}e^{U_{2\rmR}/\kBT}$. 

\begin{figure}[t]
\sidecaption[t]
\includegraphics[width=3.8in,clip]{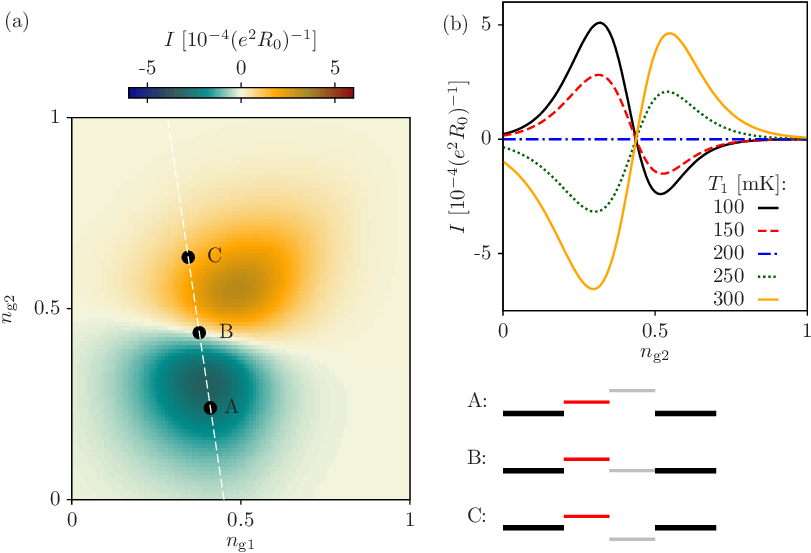}
\caption{\label{curr}\small  Transport from a hot spot. (a) Current as a function of the island control parameters, $n_{{\rm g}i}$ for an unbiased configuration $V_\rmL=V_\rmR=0$ and $T=T_2=200$~mK. The temperature of island 1 is increased at $T_1=150$~mK, with $R_\rmL=R_\rmC=R_\rmR=R_0$, $E_1=E_2=0.15$~meV, and $J=0.05$~meV. The configurations in points A, B and C are sketched on the right bottom side. (b) Cut along the white dashed line in (a) as a function of $n_{{\rm g}2}$. Island 1 is tuned such that $U_{1\rmL}=0.015$~meV is constant. Reversing the temperature gradient or the sign of $U_{2\rmR}$ changes the sign of the current.}
\end{figure}

The current generated in such a configuration with $T_1\neq T$ is plotted in Fig.~\ref{curr}(a) as a function of the control parameters $n_{{\rm g}1}$ and $n_{{\rm g}2}$. Fixing the gate voltage in island 1, the current changes sign when the gate voltage of island 2 is tuned. As depicted in the inset, the sign of the current depends on the sign of $U_{1\rmL}+U_{21}$: electrons tunneling out from the hot island into the region $i$ giving the largest $U_{1i}$ have a larger rate. The response is restricted to a region of gate voltages such that $U_{li}\lesssim\kBT$. Far from this region, the difference of the two rates is exponentially small and therefore current is suppressed.

Reversing the sign of the local temperature gradient, $T_1-T$, changes the sign of the current, cf. Fig.~\ref{curr}(b). Being an obvious statement, it has practical consequences: rather than by heating one of the islands, the effect can be experimentally detected by cooling it. This can be done locally and non-invasively by a coupled refrigerator system within nowadays technology~\cite{timofeev,partanen_quantum_2016,sinis}. 

A particularly interesting configuration is when $n_{{\rm g}2}$ is tuned such that $U_{2\rmR}=0$. This is configuration B in Fig.~\ref{curr}. Then, the rates $\Gamma_{\rmR2}=\Gamma_{2\rmR}=\kBT/(e^2R_\rmR)$ only contribute to the prefactor of the current and the second island plays no role.  In this case, $U_{1\rmL}=U_{12}$, i.e. the temperature gradient and the energy cost are the same for tunneling through the two tunneling junctions of island 1. Hence electrons in the hot island have no preferred direction to tunnel out. Thus detailed balance is symmetrically broken in the two junctions, with the overall rate through the island being the same in both directions: $\Gamma_{1\rmL}\Gamma_{21}=\Gamma_{12}\Gamma_{\rmL1}$, and the current is zero, see Eq.~(\ref{eq:curr}). 


\section{Multijunction arrays}
\begin{figure}[t]
\sidecaption[t]
\includegraphics[width=3.5in,clip]{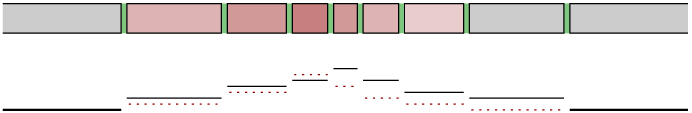}
\caption{\label{array}\small  An array of islands of different size in the presence of a local temperature increase. The charging energy of each island depends on its size, hence introducing an spacial modulation of the Coulomb gap (black solid lines). The piecewise-constant temperature distribution (red dotted lines) allows for a wider extension due to heat leaking to neighbour islands. }
\end{figure}

One can envision to extend the above results to longer arrays, cf. Fig.~\ref{array}. This relaxes the requirement to increase the temperature in a micrometer-size single island. The size of the islands can be controlled in the sample growth, which introduces a natural way to spatially modulate the Coulomb gap along the conductor. This way the need to gate the system can also be avoided. The current is then induced by the interplay of local non-equilibrium (only due to a piecewise-constant temperature distribution) and electron-electron interactions, emphasizing the mesoscopic nature of the device. 

\section{Summary}
A simple model of locally broken detailed balance giving rise to transport in an electronic conductor is presented. A system of two metallic islands, one of which experiences a different temperature, can be interpreted as a mesoscopic analogue of transport in state-dependent diffusion at a single scatterer. Dividing the conductor in an array of metallic islands allows for the local control of voltages and temperatures. The cooperative occurrance of a local temperature difference and a Coulomb gap introduces a preferred direction for tunneling electrons. This asymmetry, which determines the sign of the current, can furthermore be tuned by means of gate voltages applied to each island. Broken detailed balance is of relevance e.g. for the investigation of the Jarzynski equality and fluctuation theorems~\cite{pekola,drag,gregory,spintronics,detection,robert}, and can be detected in higher-order cummulants of the full counting statistics of Coulomb-blockade systems~\cite{philipp}.

The electron-hole asymmetry required for having a thermoelectric response is introduced by inhomogeneous energy costs in the tunneling processes. This is the case of islands with different charging energies or which are modulated by gate voltages. Thus energy-dependence of the barriers is not necessary. The involved technology is within nowadays reach~\cite{pothier1992,lotkhov,limbach,timofeev,pekola,partanen_quantum_2016} and could readily be tested in an experiment. This proposal contributes to the field of interaction-induced thermoelectric properties~\cite{coulombgap,anna} and the control of thermal flows~\cite{ruokola,transistor} in low-dimensional metallic conductors.

\begin{acknowledgement}
I acknowledge Markus B\"uttiker and Jukka P. Pekola for discussions and comments on an earlier version of the results presented here. This work was supported by the Spanish Ministerio de Econom\'ia y Competitividad via grants No. MAT2014-58241-P, No. FIS2015-74472-JIN (AEI/FEDER/UE) and RYC-2016-20778. I also thank the COST Action MP1209 ``Thermodynamics in the quantum regime". 
\end{acknowledgement}
%
%
%


\end{document}